\documentclass{aa}

\pdfoutput=1

\usepackage{amsmath}
\usepackage{bm}
\usepackage{color}
\usepackage{graphicx}
\usepackage{hyperref}
\usepackage{multirow}
\usepackage{natbib}
\usepackage{flushend}
\usepackage{txfonts}

\usepackage{natbib}
\bibpunct{(}{)}{;}{a}{}{,} 

\hypersetup{colorlinks,
  linkcolor=blue,
  filecolor=blue,
  urlcolor=blue,
  citecolor=blue}

\def\lcdm{$\Lambda$CDM}
\def\planck{{\it Planck}}
\def\plck{PLCK~G004.5$-$19.5}
\def\pygmos{\texttt{PyGMOS}}
\def\sz{Sunyaev-Zel'dovich}
\def\xmm{{\it XMM-Newton}}
\def\lenstool{{\sc LensTool}}
\def\emass{$M_E=2.45_{-0.47}^{+0.45}\times10^{14}\,M_\odot$}
\def\mass{$M_{500}^{SL}=4.0_{-1.0}^{+2.1}\times10^{14}M_\odot$}
\def\mslx{$M_{500}^{SL+X}=6.7_{-1.3}^{+2.6}\times10^{14}M_\odot$}
\def\tma{\tablefootmark{a}}
\def\tmb{\tablefootmark{b}}
\def\tmc{\tablefootmark{c}}
\def\tmd{\tablefootmark{d}}
\def\tme{\tablefootmark{e}}

\def\tmab{$^{a,b}$}

\def\aar{Astron.\ Astrophys.\ Rev.}
\def\aspc{ASPC}
\def\casp{Comm.\ Astrophys.\ Space\ Phys.}
\def\nature{Nature}
\def\njph{New~J.~Phys.}

\begin{document}

\title{Strong Lensing Analysis of \plck,
       a Planck-Discovered Cluster Hosting a Radio Relic at $z=0.52$
       \thanks{Based on observations obtained at the Gemini Observatory,
               which is operated by the Association of Universities for Research
               in Astronomy, Inc., under a cooperative agreement with the NSF on
               behalf of the Gemini partnership: the National Science Foundation
               (United States), the Science and Technology Facilities Council
               (United Kingdom), the National Research Council (Canada), CONICYT
               (Chile), the Australian Research Council (Australia),
               Minist\'{e}rio da Ci\^{e}ncia, Tecnologia e Inova\c{c}\~{a}o
               (Brazil) and Ministerio de Ciencia, Tecnolog\'{i}a e
               Innovaci\'{o}n Productiva (Argentina)}}

\author{Crist\'obal~Sif\'on,\inst{1,7}
        Felipe~Menanteau,\inst{2,3,7}
        John~P.~Hughes,\inst{4,7}
        Mauricio~Carrasco\inst{5,6}
        \and L.~Felipe~Barrientos\inst{5}}

\institute{Leiden Observatory,
           Leiden University,
           PO Box 9513, NL-2300 RA Leiden,
           Netherlands
           \and
           National Center for Supercomputing Applications,
           University of Illinois at Urbana-Champaign,
           1205 W. Clark St, Urbana, IL 61801, USA
           \and
           University of Illinois at Urbana-Champaign,
           Department of Astronomy,
           1002 W. Green Street, Urbana, IL 61801, USA
           \and
           Rutgers University,
           Department of Physics \& Astronomy,
           136 Frelinghuysen Rd, Piscataway, NJ 08854, USA
           \and
           Departamento de Astronom\'ia y Astrof\'isica,
           Facultad de F\'isica,
           Pontificia Universidad Cat\'olica de Chile,
           Casilla 306, Santiago 22, Chile
           \and
           Zentrum f\"ur Astronomie,
           Institut f\"ur Theoretische Astrophysik,
           Philosophenweg 12, 69120 Heidelberg, Germany
           }

\abstract
 {The recent discovery of a large number of galaxy clusters using the \sz\ (SZ) effect has opened a
  new  era on the study of the most massive clusters in the Universe. Multi-wavelength analyses are
  required to understand the properties of these new sets of clusters, which are a sensitive probe
  of cosmology.}
 {We aim at a multi-wavelength characterization of \plck, one of the most massive X-ray validated
  SZ effect--selected galaxy clusters discovered by the \planck\ satellite.}
 {We have observed \plck\ with GMOS on the 8.1m-Gemini South Telescope for optical imaging and
  spectroscopy, and performed a strong lensing analysis. We also searched for associated radio
  emission in published catalogs.}
 {An analysis of the optical images confirms that this is a massive cluster, with a dominant
  central galaxy (the BCG) and an accompanying red sequence of galaxies, plus a 
  $14\arcsec$-long strong lensing arc. Longslit spectroscopy of 6 cluster members
  shows that the cluster is at $z=0.516\pm0.002$. We also targeted the strongly lensed arc, and 
  found $z_{\rm arc}=1.601$. We use \lenstool\ to carry out a strong lensing analysis,
  from which we measure a median Einstein radius $\theta_E(z_s=1.6)\simeq30\arcsec$ and estimate an
  enclosed mass \emass. By extrapolating an NFW profile we find a total mass \mass. Including a
  constraint on the mass from previous X-ray observations yields a slightly higher mass, \mslx,
  marginally consistent with the value from strong lensing alone. High-resolution radio images from
  the TIFR GMRT Sky Survey at 150~MHz reveal that \plck\ hosts a powerful radio relic on scales
  $\lesssim500$ kpc. Emission at the same location is also detected in low resolution images at
  843~MHz and 1.4~GHz. This is one of the higher redshift radio relics known to date.}
 {}


\titlerunning{Strong Lensing analysis of \plck}
\authorrunning{C.~Sif\'on et al.}

\maketitle

\footnotetext[7]{Visiting astronomer, Gemini South Observatory}

\section{Introduction}

\begin{figure*}
 \centerline{\includegraphics[width=\textwidth]{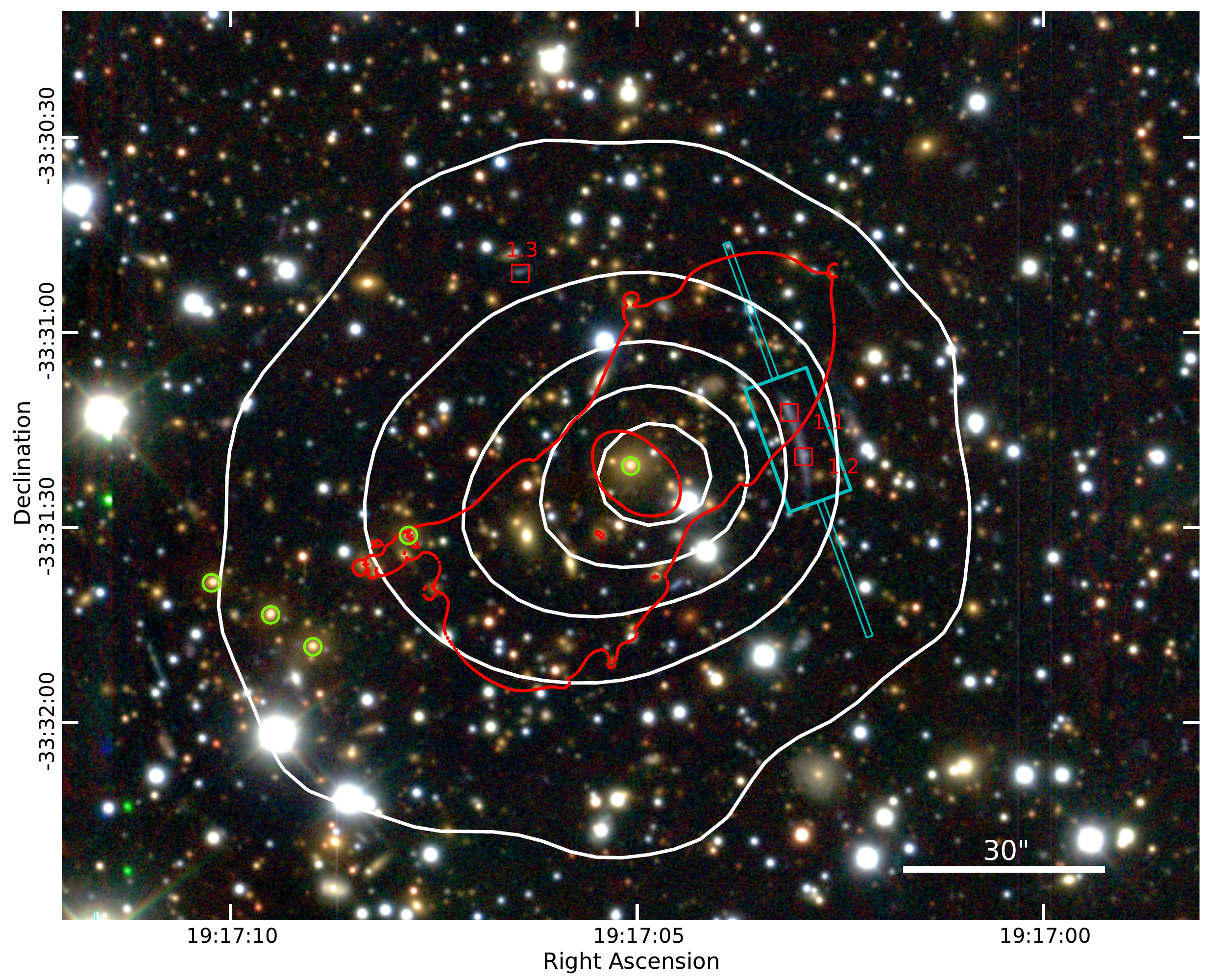}}
\caption{GMOS {\it gri} pseudo-color image of the central region of \plck. North is up, east is
left. X-ray surface brightness contours from \xmm\ are overlaid in white. Spectroscopic cluster
members are marked by green circles; only 5 out of 6 are visible in the shown region, the sixth
member is $\sim760$ kpc to the E-SE of the BCG. Red squares mark the position of the 3 confirmed 
multiple images, while we show in red the critical curve for $z_s=1.6$. The thin cyan box shows the 
slit used to get the spectrum of the arc ($1\arcsec$ across) ; the wide cyan box shows the region
zoomed-in in the left panel of Figure \ref{f:arc}. The thick white line in the bottom right shows a 
$30\arcsec$ scale, corresponding to 188~kpc at $z=0.516$.}
\label{f:gri}
\end{figure*}

In the last few years, the \sz\ (SZ) effect has proven to be an effective method to find massive
galaxy clusters at all redshifts, with results from the Atacama Cosmology Telescope \citep[ACT,
e.g.,][]{marriage11,hasselfield13}, the South Pole Telescope \citep[SPT,
e.g.,][]{williamson11,reichardt13} and the \planck\ satellite \citep[e.g.,][]{planck_esz,planck_psz}
already yielding a few hundred newly discovered clusters up to $z\sim1.4$. The SZ effect is a
distortion in the Cosmic Microwave Background (CMB) spectrum in the direction of galaxy clusters
caused by inverse Compton scattering of CMB photons by the hot electrons in the intracluster gas
\citep{sunyaev72}. Multi-wavelength follow-up observations of SZ-selected clusters have confirmed
the unique potential of the SZ effect for detecting the most massive clusters in the Universe
\citep[e.g.,][]{benson13,sifon13}, with the SZ-discovered El Gordo and SPT-CL~J2344$-$4243 being
two of the most extreme galaxy clusters ever known \citep{gordo,mcdonald12}. As expected, many of
these clusters display strong lensing features \citep{menanteau10a}, a good indication that these
are very massive systems.

Observations of these strongly lensed background galaxies offer one of the most robust ways of
constraining the mass of a cluster, providing a direct measure of the mass within the Einstein
radius \citep[see][for a recent review]{kneib11}. In combination with other probes (such as X-rays
and weak lensing), strong lensing analyses have provided some of the most complete mass distribution
models for galaxy clusters, even allowing for the determination of the 3-dimensional configuration
in some cases \citep[e.g.,][]{morandi10, limousin13}.

Here, we present a multi-wavelength analysis of \plck, one of the most massive, hot and X-ray 
luminous galaxy clusters discovered by the \planck\ satellite via the SZ effect and validated with 
\xmm\ X-ray observations \citep{planck_xmm}. We perform a strong lensing analysis from 
optical imaging and spectroscopy, and show from archival radio imaging that \plck\ hosts a 
powerful radio relic.

All uncertainties are quoted at the 68.3\% ($1\sigma$) confidence level. We assume a flat \lcdm\
cosmology with $\Omega_M=0.3$ and $H_0=70\,\mathrm{km\,s^{-1}\,Mpc^{-1}}$. Total masses, X-ray and 
SZ measurements are reported within a radius $r_{500}$, which encloses a mean density 500 
times the critical density of the Universe at the corresponding redshift. All quantities reported by
\cite{planck_xmm} (reproduced in Sec.\ \ref{s:previous}) have been corrected to the spectroscopic 
redshift $z=0.516$. All magnitudes are in the AB system.

\section{Observations and Data Analysis}

\subsection{SZ and X-ray Data}\label{s:previous}

\plck\ was discovered through its SZ effect by the \planck\ satellite. With a signal-to-noise ratio
(S/N) of 5.9 in the Early Science release, it is just below the S/N threshold of 6.0 set for the 
\planck\ Early SZ sample \citep{planck_esz}\footnote{\plck\ has been included in the new \planck\ 
SZ catalog \citep{planck_psz} with a S/N of 6.15.}. Despite this relatively low S/N, it has a strong
integrated SZ signal, $Y_{500} = (1.90\pm0.19)\times10^{-4}\,\mathrm{Mpc^2}$, where $Y \equiv \int
y\,d\Omega$. Here, $y$ is the usual Compton parameter and the integral is over the solid angle of
the cluster. We use the $Y-M$ scaling relation of \cite{planck_scaling} to estimate a mass
$M_{500}^{SZ}=(10.4\pm0.7)\times10^{14}\,M_\odot$.

\plck\ was subsequently validated using \xmm\ \citep{planck_xmm}, which confirmed that it is an
extended X-ray source. Moreover, the observed energy of the Fe K emission line allowed a redshift
determination $z_{\rm Fe}=0.54$, making it the highest-redshift cluster of the initial
\planck--\xmm\ validation program. The X-ray analysis of \cite{planck_xmm} proves that \plck\ is a
hot, massive cluster, with an X-ray luminosity\footnote{The uncertainties in the X-ray values from
\cite{planck_xmm} do not include systematic errors and have been dropped when negligible.} (in the
[0.1-2.4] keV band) of $L_X=1.6\times10^{45}\,\mathrm{erg\,s^{-1}}$, an integrated temperature
$kT_X=10.2\pm0.5$ keV and a gas mass $M_{\rm gas}=1.3\times10^{14}\,M_\odot$. Combined, the latter
two give a pseudo-Compton parameter
$Y_X \equiv kT_X M_{\rm gas} = \left(13.3\pm0.9\right)\times10^{14}M_\odot$ keV.
With this latter value, \cite{planck_xmm} estimate a total mass 
$M_{500}^X=\left(9.6\pm0.5\right)\times10^{14}\,M_\odot$.

\subsection{Optical Imaging}\label{s:img}

\plck\ was observed on UT 2012 July 19 with the {\it gri} filters with GMOS on the Gemini-South
Telescope (ObsID:GS-2012A-C-1, PI:Menanteau), with exposure times of $8\times60$~s, $8\times90$~s
and $8\times150$~s respectively. Observations were performed with photometric conditions and seeing
$\sim0.\!\arcsec6$. Images were coadded using SWarp \citep{swarp} and photometry was performed using
SExtractor \citep{sextractor} in dual mode, using the {\it i}-band for detection. Figure \ref{f:gri}
shows the combined {\it gri} image\footnote{Created with {\sc stiff} \citep{stiff}.} of \plck, which
shows clearly that there is an overdensity of red elliptical galaxies with a central dominant
Brightest Cluster Galaxy (BCG) close to the X-ray peak. Figure \ref{f:gri} also reveals the presence
of several strong lensing features, most notably a giant arc to the West of the BCG, roughly
$14\arcsec$ long.

Each galaxy is assigned a photometric redshift by fitting Spectral Energy Distributions (SEDs) to
the {\it gri} photometry using the BPZ code \citep{benitez00} including correction for galactic
extinction as described in \cite{menanteau10a,menanteau10b}. Typical uncertainties are $\delta
z/(1+z)\simeq0.09$. The photometric redshift of the cluster, $z_{\rm phot}=0.51\pm0.02$, was
estimated as in \cite{menanteau10a,menanteau10b} and is consistent with the spectroscopic redshift
(Sec.~\ref{s:spec}). We consider as cluster members all galaxies within $\Delta z=0.03(1+z_0) =
0.045$ of $z_0=0.51$ and brighter than $m^\star+2\simeq22.9$ in the {\it i}-band, for a total 222
photometrically-selected members. (Here $m^\star$ is the characteristic luminosity of the 
\cite{schechter76} function as found by \cite{blanton03}, passively evolved to $z_0$.\footnote{For 
reference, the BCG has a luminosity $L=9.5L^\star$.}) Selecting galaxies from a color-magnitude 
diagram instead or imposing a brighter membership cut have no influence on the results.

\subsection{Optical Spectroscopy}\label{s:spec}

\begin{table}
\begin{center}
\caption{Spectroscopically confirmed cluster members.}
\label{t:redshifts}
\begin{tabular}{l c c c c}
\hline\hline
ID &     RA     &     Dec    & $i$ mag.\tma & Redshift\tmb \\
   & (hh:mm:ss) & (dd:mm:ss) &   (AB mag)   &              \\[0.5ex]
\hline
1\tmc  & 19:17:05.08 & $-$33:31:20.6 & 18.47 & $0.5199\pm0.0005$ \\
2      & 19:17:07.80 & $-$33:31:31.2 & 19.74 & $0.5126\pm0.0005$ \\
3      & 19:17:08.98 & $-$33:31:48.4 & 19.16 & $0.5074\pm0.0004$ \\
4      & 19:17:09.49 & $-$33:31:43.5 & 19.70 & $0.5150\pm0.0003$ \\
5      & 19:17:10.20 & $-$33:31:38.5 & 19.83 & $0.5176\pm0.0003$ \\
6      & 19:17:14.40 & $-$33:31:57.5 & 20.48 & $0.5187\pm0.0002$ \\
\hline
\end{tabular}
\end{center}
\tablefoottext{a}{\texttt{MAG\_AUTO} from SExtractor.}\\
\tablefoottext{b}{Errors as given by RVSAO.}\\
\tablefoottext{c}{Brightest Cluster Galaxy.}
\end{table}

We performed longslit spectroscopy of \plck\ on UT 2012 July 20 with GMOS, with $0.\!\arcsec75$-wide
slits with three pointings, two aimed at confirming cluster members and one targeting the most
prominent strongly lensed background galaxy. The data were reduced using
\pygmos\footnote{\url{http://www.strw.leidenuniv.nl/\~sifon/pygmos/}} \citep{sifon13}, with an
average wavelength calibration root-mean-square (rms) uncertainty of 0.4\AA. Redshifts were measured
by cross-correlating the spectra with Sloan Digital Sky Survey \citep{sdss} template spectra using
the IRAF package RVSAO \citep{rvsao}. The six confirmed cluster members are listed in Table
\ref{t:redshifts} and are shown in Figure \ref{f:gri} by green circles. They are all red, passive
elliptical galaxies and have a rest-frame velocity dispersion $\sigma\sim860\,\mathrm{km\,s^{-1}}$
(which is likely not representative of the cluster velocity dispersion). The median redshift of
these 6 members, $z=0.516\pm0.002$, is adopted as the cluster redshift (with uncertainties given by
$\sigma\sqrt{\pi/2N}$).

\begin{figure*}
\centerline{\includegraphics[width=\textwidth]{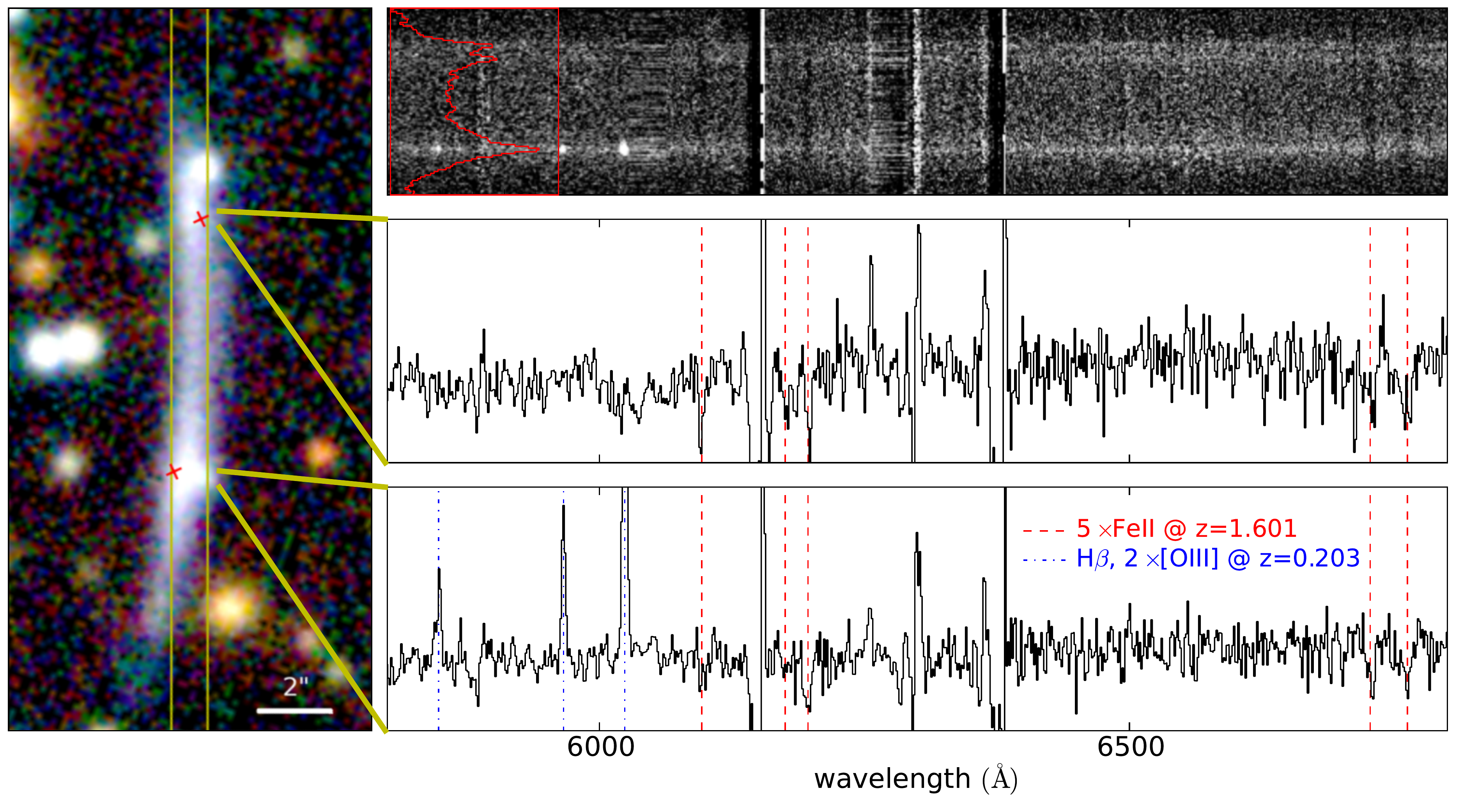}}
\caption{Strong lensing giant arc. The left panel shows a $10\arcsec\times20\arcsec$ close-up {\it 
gri} image 
of the arc (cyan box in Figure \ref{f:gri}), with the red crosses marking the location where 
\lenstool\ predicts the images to be. The thin yellow lines outline the position of the slit and 
the thick yellow lines mark the approximate locations of the knots from where the 1D spectra are 
shown. The right panels show the arc spectrum in the wavelength range $5800\AA-6800\AA$. The top 
right panel shows the GMOS 2d spectrum. The image is 105 pixels, corresponding to $15.\!\arcsec3$, 
from top to bottom. The red histogram (inset) shows the total counts in each row over the shown 
spectral range, after an iterative $3\sigma$-clipping to remove bad pixels and emission lines. This 
highlights the decrease in brightness (and the significance of the continuum) between the two 
images. The middle and bottom panels show, respectively, the 1d spectra of the northern (source 1.1) 
and southern (source 1.2) peaks seen in the lensed arc, each marked by a yellow ``wedge'' in the 
left panel. In these, the red dashed lines mark the 5 FeII absorption lines at $z=1.601$ and the 
blue dash-dotted lines mark the emission lines from a foreground galaxy at $z=0.203$, only seen in 
the south spectrum. The vertical axes in these two panels are in arbitrary units.}
\label{f:arc}
\end{figure*}

The left panel of Figure \ref{f:arc} shows a zoomed-in view of the brightest lensed galaxy. Two
brightness peaks can be identified, which we interpret as two blended strong lensing images of a
single source (see Sec.\ \ref{s:sl}). The top-right panel shows the 2d spectrum along the arc, where
a faint continuum can be distinguished between the north and south images. The red inset histogram
shows the normalized counts for each row over the spectral range shown, after an iterative
$3\sigma$-clipping rejection so that bad pixels and emission lines are not included in the counts.
This histogram shows that the decrease in brightness is significant between the two peaks but that
this region is, in turn, still detected at high significance. The middle- and bottom-right panels
show the 1d spectra of the two brightness peaks. Both spectra clearly show 5 redshifted FeII
absorption lines with rest-frame wavelengths 2344.2, 2374.5, 2382.8, 2586.6 and 2600.2 \AA. The
median redshift of these 5 pairs of lines is $z_{\rm arc}=1.6008\pm0.0002$. The bottom spectrum also
shows three emission lines (seen in the 2d spectrum as well), which correspond to H$\beta$ and
[OIII]$\lambda\lambda$4958,5007\AA\ from a foreground compact star-forming galaxy at $z=0.203$, for
which H$\alpha$ emission is also observed but not shown in Figure \ref{f:arc}. A small, bright, 
blue blob is indeed seen overlapping with the south knot (just West of the latter), which we 
interpret as this foreground galaxy.

\section{Strong Lensing Analysis}\label{s:sl}

\subsection{Strong Lensing Model}

\begin{figure*}
 \centerline{\includegraphics[width=\textwidth]{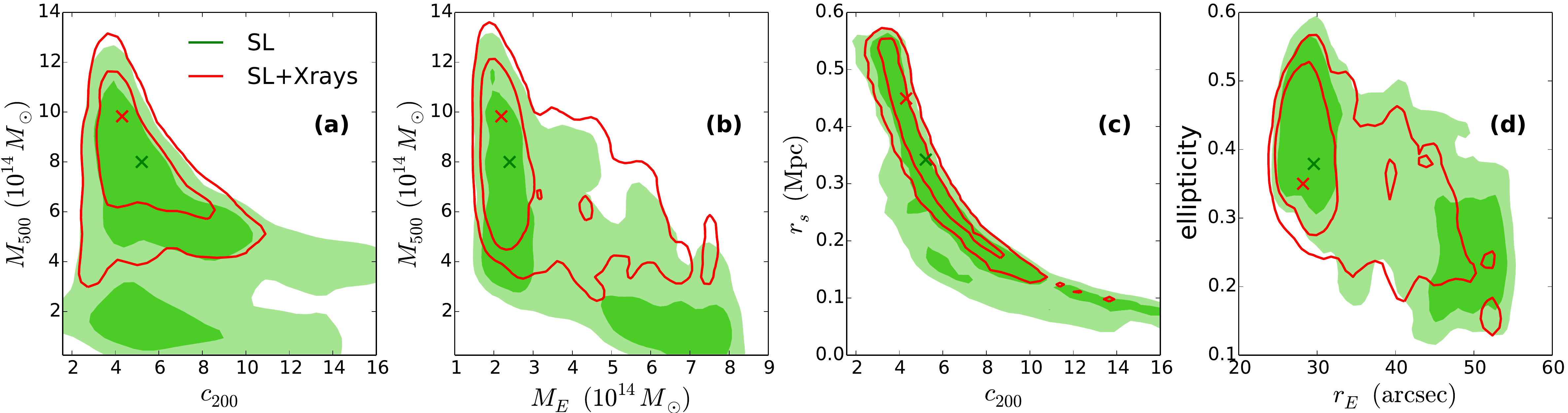}}
\caption{Joint 2D posterior distributions of $c_{200}$ and $M_{500}$ (panel {\it a}), $M_E$ and 
$M_{500}$ (panel {\it b}), $c_{200}$ and $r_s$ (panel {\it c}), and $r_E$ and $\theta$ (panel {\it 
d}). Contours are at the 68\% and 95\% levels. Filled green contours show constraints from strong 
lensing alone and red contours show the constraints when $M_{500}^X$ is included as an independent 
constraint. Crosses show the corresponding maximum likelihood estimates.}
\label{f:contours}
\end{figure*}

The strong lensing analysis was performed using the Markov Chain Monte Carlo (MCMC) code \lenstool\
\citep{kneib93,jullo07}, as follows. The cluster is modelled with an ellipsoidal Navarro-Frenk-White
\citep[NFW,][]{nfw95} profile for the main halo, plus a truncated Pseudo-Isothermal Elliptical Mass
Distribution \citep[PIEMD,][]{kassiola93,kneib96} with a constant mass-to-light ratio for the 222
brightest cluster members (see Sec.~\ref{s:img}). A PIEMD halo is modelled by three parameters: the
core radius, $r_{\rm core}$, the size of the halo (the cut-off radius), $r_{\rm cut}$, and the
velocity dispersion, $\sigma_0$, which scale with galaxy luminosity as \citep{jullo07}:
\begin{subequations}\label{eq:piemd}
\begin{align}
 r_{\rm core} &= r^\star_{\rm core} \left(L/L^{\star}\right)^{1/2} \\
  r_{\rm cut} &= r^\star_{\rm cut} \left(L/L^{\star}\right)^{1/2} \\
     \sigma_0 &= \sigma^\star_0 \left(L/L^{\star}\right)^{1/4},
\end{align}
\end{subequations}
where $L^\star=6.6\times10^{10}\,L_\odot$. The total mass of the galaxy is then given by
\begin{equation}
 M = \left(\pi/G\right) \left(\sigma^\star_0\right)^2 r^\star_{\rm cut} \left(L/L^{\star}\right).
\end{equation}
We fix $r^\star_{\rm core}=0.3$ kpc, and $r^\star_{\rm cut}$ and $\sigma^{\star}_0$ are free
parameters. The center of the NFW halo is fixed to the peak of the X-ray emission 
\citep[RA=19:17:04.6, Dec=$-$33:31:21.9;][]{planck_xmm}. Therefore the mass model has six free 
parameters: four for the main NFW halo and two for the PIEMD halos.

As can be seen in the red inset histogram of Figure \ref{f:arc}, there is a decrease in brightness 
in the middle of the arc in between two prominent brightness peaks. We interpret this as the 
merging of two images of the background galaxy and use this double-imaged arc with $z_{\rm 
arc}=1.6$ as a constraint for the lens model, and identify a third image of the same source to the 
North-East of the BCG (labelled 1.3 in Figure \ref{f:gri}). The positions and photometry of these 
three images are listed in Table \ref{t:images}.

The total mass model is therefore optimized using the 222 brightest members (including the six
spectroscopic members) and the three images for the background galaxy at $z=1.601$. We adopt a
positional uncertainty $\Delta{\bf x}=1.\!\arcsec4$ for the multiple images. The goodness-of-fit for
the best model is $\chi^2_{\rm red}/{\rm d.o.f.} = 0.15$, with a rms error on the image positions of 
$0.\!\arcsec22$. The total mass distribution is moderately elongated along the plane of the sky, 
approximately aligned with the light distribution. The best-fit values for the six free parameters 
plus the posterior masses and radii are listed in Table \ref{t:model} (see also Sec.\ 
\ref{s:extra}).

Following \cite{meneghetti11}, the Einstein radius is estimated as the median distance of the
tangential critical curves to the cluster center. We find
 $\theta_E(z_s=1.6)=30.\!\arcsec3_{-3.9}^{+1.4}$,
corresponding to a physical distance $r_E\simeq190$ kpc. Assuming a symmetric lens, the mass inside
this region is \emass. Integrating the 3-dimensional NFW profile for the main halo, we obtain \mass.
The corresponding radius, $r_{500}^{SL}=0.93_{-0.08}^{+0.16}$ Mpc, is estimated from $M_{500}$ assuming
a spherical cluster. We note that the values at $r_{500}$ are an extrapolation of the strong lensing 
information.

Recently, \cite{zitrin12} derived a representative distribution of Einstein radii from a sample of
$\sim10,000$ clusters from the SDSS optically-selected sample of \cite{hao10}. They found a
log-normal Einstein radius distribution with mean and standard deviation
 $\langle\log(\theta_E^{\rm eq}/{\rm arcsec})\rangle=0.73\pm0.32$
for background sources at $z_s\sim2$. For comparison to \cite{zitrin12} and others, we estimate the 
equivalent Einstein radius to be
 $\theta_E^{\rm eq}(z_s=1.6)\simeq25\arcsec$.
\plck\ is a $2\sigma$ outlier from this mean relation; therefore it can be said to be within the 5\% 
strongest lensing clusters in the Universe.

\begin{table}
\begin{center}
\caption{Images of the strongly lensed galaxy.}
\label{t:images}
\begin{tabular}{c c c c c}
\hline\hline
 Source & RA & Dec & $r$ mag.\tma & $g-r$\tmb \\
  & (hh:mm:ss) & (dd:mm:ss) & (AB mag) & (AB mag) \\
\hline
1.1 & 19:17:03.14 & $-$33:31:12.5 & $21.42\pm0.01$ & $0.40\pm0.03$ \\
1.2 & 19:17:02.97 & $-$33:31:19.0 & $21.42\pm0.01$ & $0.39\pm0.02$ \\
1.3 & 10:17:06.45 & $-$33:30:50.8 & $23.75\pm0.03$ & $0.35\pm0.05$ \\
\hline
\end{tabular}
\end{center}
\tablefoottext{a}{\texttt{MAG\_ISO} from SExtractor.} \\
\tablefoottext{b}{Difference of \texttt{MAG\_APER}'s from SExtractor.}
\end{table}

\begin{table}
\begin{center}
\caption{Marginalized posterior estimates of the strong lensing model with and without the X-ray
mass constraint.}
\label{t:model}
\begin{tabular}{l c r@{}l r@{}l l}
\hline\hline
Parameter\tma & Symbol & \multicolumn{2}{c}{SL} & \multicolumn{2}{c}{SL+X} & units\\[0.5ex]
\hline
\multicolumn{7}{c}{Main NFW Halo} \\[0.10cm]
Ellipticity         & $e$         & $0.40$ & $_{-0.09}^{+0.07}$ & $0.37$ & $_{-0.07}^{+0.08}$ &     
\\[0.12cm]
Position Angle\tmb  & $\theta$    &   $52$ & $_{-3}^{+7}$       &   $53$ & $_{-1}^{+2}$       & deg 
\\[0.12cm]
Scale radius        & $r_s$       & $0.10$ & $_{-0.04}^{+0.17}$ & $0.39$ & $_{-0.08}^{+0.07}$ & Mpc 
\\[0.12cm]
Concentration\tmc   & $c_{200}$   &  $4.0$ & $_{-0.8}^{+5.0}$   &  $4.1$ & $_{-0.6}^{+1.8}$   &     
\\[0.12cm]
\hline
\multicolumn{7}{c}{PIEMD Halos} \\[0.10cm]
Cut-off radius      & $r^\star_{\rm cut}$ &  $47$ & $_{-20}^{+13}$ &  $25$ & $_{-14}^{+28}$ & kpc 
\\[0.12cm]
Velocity dispersion & $\sigma^\star_0$    & $225$ & $_{-23}^{+42}$ & $106$ & $_{-53}^{+37}$ &
$\mathrm{km\,s^{-1}}$ \\[0.12cm]
\hline
\multicolumn{7}{c}{Derived Parameters} \\[0.10cm]
Einstein Mass   & $M_E$ & 2.45 & $_{-0.47}^{+0.45}$ & 2.46 & $_{-0.59}^{+0.31}$ & $10^{14}M_\odot$ 
\\[0.12cm]
Einstein Radius & $r_E$ & 30.3 &   $_{-3.9}^{+1.4}$ & 30.0 &   $_{-3.5}^{+0.6}$ & arcsec \\[0.12cm]
Total Mass  & $M_{500}$ & 4.0  &   $_{-1.0}^{+2.1}$ &  6.7 &   $_{-1.3}^{+2.6}$ & $10^{14}M_\odot$ 
\\[0.12cm]
Radius      & $r_{500}$ & 0.93 & $_{-0.08}^{+0.16}$ & 1.10 & $_{-0.07}^{+0.14}$ & Mpc \\[0.12cm]
\hline
\end{tabular}
\end{center}
\tablefoottext{a}{All parameters have uniform priors.}\\
\tablefoottext{b}{Position angle West of North.}\\
\tablefoottext{c}{The concentration is defined as $c_{200}=r_{200}/r_s$.}
\end{table}

\subsection{External Constraints}\label{s:extra}

We run \lenstool\ again including a prior in the mass, from the X-ray mass estimated by 
\cite{planck_xmm} as implemented by \cite{verdugo11}. As mentioned in Sec.\ \ref{s:previous}, 
however, the reported uncertainties are unrealistically small. As a more realistic estimate, we take 
the intrinsic scatter in the latest $Y_X-M$ relation by \cite{mahdavi13} of 22\%, measured by 
combining weak lensing and X-ray observations. Thus the additional constraint in the total mass is 
the following Gaussian prior:
\begin{equation}\label{eq:Mx}
 M_{500}^X = (9.6\pm2.1)\times10^{14}\,M_\odot
\end{equation}
measured at $r_{500}^X=1245$ kpc.\footnote{Note that in \lenstool\ the X-ray constraint to the
strong lensing model is given as a fixed mass $M$ at a fixed radius $r$ (with a mass uncertainty),
not explicitly as the mass at a given overdensity.} The same excersise for the SZ mass, assuming an 
uncertainty of 18\% corresponding to the central value of the intrinsic scatter in the $Y_{SZ}-M$ 
measured by \cite{sifon13} using dynamical masses and SZ measurements from ACT, gives
\begin{equation}\label{eq:Msz}
 M_{500}^{SZ} = (10.4\pm1.9)\times10^{14}\,M_\odot
\end{equation}
We only use Equation \ref{eq:Mx} because both measurements are very similar and because they are 
both measured at the same radius, determined from the X-ray scaling relation \citep{planck_xmm} and
are therefore not independent. The posterior distributions are shown for various combinations of 
parameters for the two different models in Figure \ref{f:contours}, highlighting degeneracies in 
the strong lensing model.

The X-ray constraint pushes the mass to a higher value which is marginally consistent with the 
strong lensing only (SL) model. Notably, the SL model allows for a low-$M_{500}$, 
high-$M_E$ (through a high $r_E$), high-concentration and low-ellipticity solution which is 
marginally excluded by the model including the X-ray constraint (SL+X). The marginalized 
posterior mass is \mslx. Although the contours are broader in the SL model, the maximum likelihood 
estimate (MLE) and marginalized 68\% range of $M_E$ (and $r_E$) are mostly unaffected by the 
inclusion of the X-ray constraint, with a posterior estimate 
$M_E^{SL+X}=2.46_{-0.59}^{+0.31}\times10^{14}M_\odot$. This is expected, since $r_E$ is directly 
constrained by the strongly lensed images, independently of the mass profile of the cluster.

\section{Radio Emission}\label{s:radio}

\begin{figure}
\centerline{\includegraphics[width=3.5in]{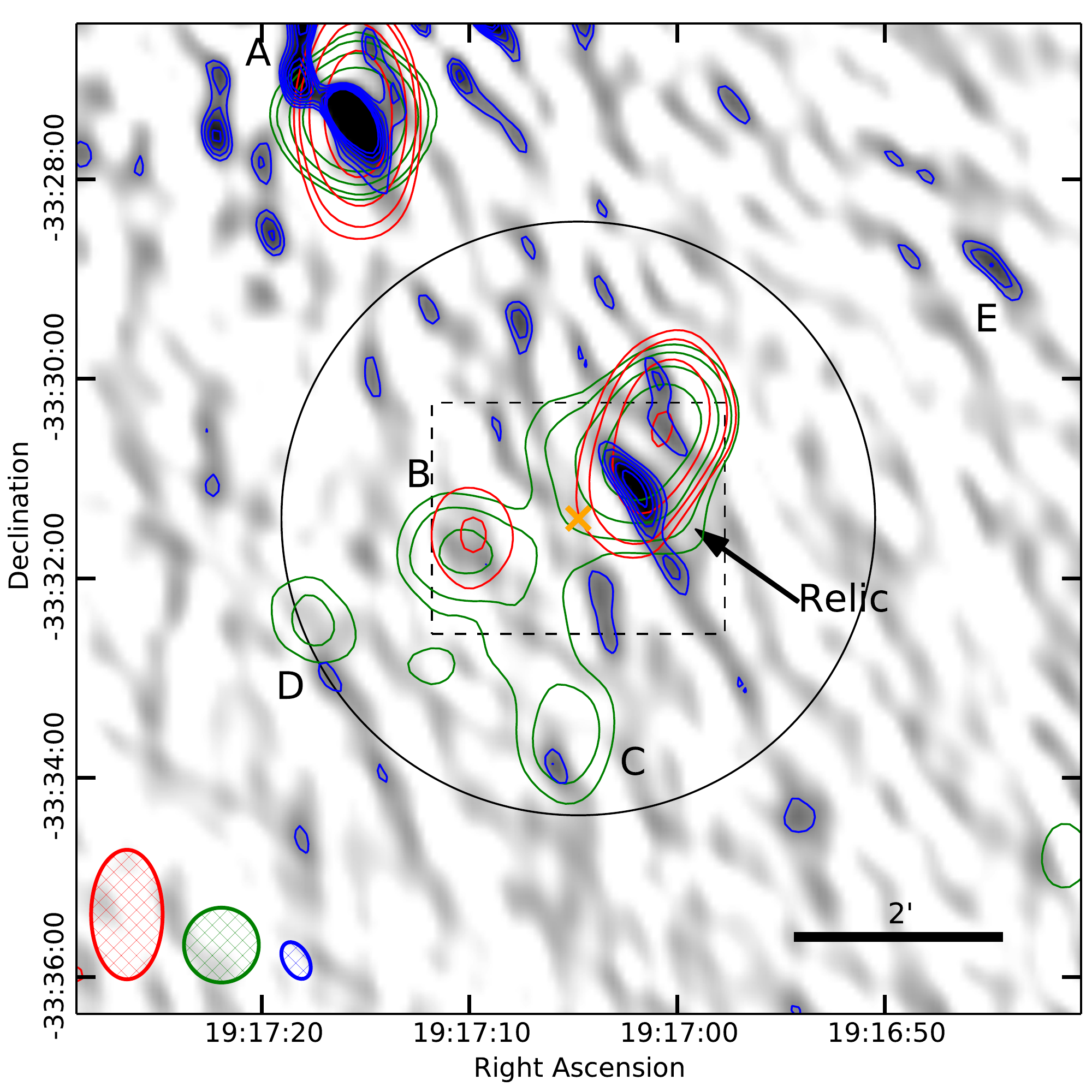}}
\caption{$10\arcmin\times10\arcmin$ TGSS 150~MHz intensity map around \plck, with contours in blue. 
Contours are shown at 3, 5, 7 and 15 times $\sigma$, where $\sigma$ is the background rms. NVSS 
1.4 GHz and SUMSS 843 MHz contours are shown in green and red, respectively. Both contour sets are 
in units of 3, 5, 10 and 20 times each $\sigma$. The orange cross shows the position of the 
BCG. The dashed black rectangle is the region shown in Figure \ref{f:gri} and the black circle 
marks $r_{500}^{SL+X}=1.12$ Mpc. The black bar in the bottom right marks a scale of 2\arcmin. The
SUMSS, NVSS and TGSS beams are shown from left to right, respectively, in the bottom left corner 
(hatched ellipses).}
\label{f:radio}
\end{figure}

\begin{table*}
\begin{center}
\caption{Radio Relic}
\label{t:relic}
\begin{tabular}{l r@{}l r@{}l r c r@{$\,\pm\,$}l c r@{$\,\pm\,$}l}
\hline\hline
Source Name & \multicolumn{2}{c}{RA\tma} & \multicolumn{2}{c}{Dec\tma} & \multicolumn{1}{c}{Freq.} 
& Beam & \multicolumn{2}{c}{$F_\nu$\tmb} & Size\tmc & 
\multicolumn{2}{c}{P.A.\tmd} \\
    & \multicolumn{2}{c}{(hh:mm:ss)} & \multicolumn{2}{c}{(dd:mm:ss)} & (MHz) & 
($\arcsec\times\arcsec$) & \multicolumn{2}{c}{(mJy)} & 
($\arcsec\times\arcsec$) & \multicolumn{2}{c}{(deg)} \\[0.5ex]
\hline
GMRT173\_01\tme        & 19:17:01.94 &     & $-$33:31:12.6 &      &  150 & $24\times15$ &
  382 & 80 & $84\times19$ &  28 & 5 \\
SUMSS~J191701$-$333033 & 19:17:01.50 &(16) & $-$33:30:33.7 &(2.5) &  843 & $51\times43$ &
   51 &  4 & $70\times62$ & 136 & 3 \\
NVSS~1917101$-$333035  & 19:17:01.75 &(08) & $-$33:30:35.6 &(1.3) & 1400 & $45\times45$ &
   37 &  2 & $74\times50$ & 146 & 1 \\
\hline
\end{tabular}
\end{center}
\tablefoottext{a}{Nominal uncertainties on the last two digits in parentheses.}\\
\tablefoottext{b}{Integrated Flux.}\\
\tablefoottext{c}{Major and minor axes.}\\
\tablefoottext{d}{Position angle West of North.}\\
\tablefoottext{e}{Position uncertainties from TGSS are $\sim4\arcsec$.}\\
\end{table*}

\begin{table*}
\begin{center}
\caption{Other Sources in the Radio Images.}
\label{t:radio}
\begin{tabular}{c l r@{}l r@{}l r@{$\,\pm\,$}l c c}
\hline\hline
Source & Name & \multicolumn{2}{c}{RA} & \multicolumn{2}{c}{Dec} & \multicolumn{2}{c}{$F_\nu$} & 
Size & P.A. \\
 &  & \multicolumn{2}{c}{(hh:mm:ss)} & \multicolumn{2}{c}{(dd:mm:ss)} & \multicolumn{2}{c}{(mJy)} & 
($\arcsec\times\arcsec$) & (deg) 
\\[0.5ex]
\hline
\multirow{3}{*}{A} &       GMRT243\_01\tmab & 19:17:16.71 &      & $-$33:27:11.8 &       &
 540 &  75 & $39\times19$ &  43 \\
                   &  NVSS~J191715$-$332722 & 19:17:15.89 & (05) & $-$33:27:22.2 & (0.7) &
35.7 & 0.5 & $47\times45$ & 176 \\
                   & SUMSS~J191715$-$332720 & 19:17:15.75 & (11) & $-$33:27:20.6 & (2.0) &
71.5 & 2.7 & $73\times49$ & 165 \\[0.15cm]
\multirow{2}{*}{B} &  NVSS~J191710$-$333144 & 19:17:10.90 & (28) & $-$33:31:44.7 & (3.3) &
11.2 & 0.8 & $74\times55$ & 102 \\
               & SUMSS~J191710$-$333139\tmc & 19:17:10.55 & (37) & $-$33:31:39.0 & (4.6) &
 8.8 & 1.0 & $57\times52$ &  3  \\[0.15cm]
C                  &  NVSS~J191705$-$333333 & 19:17:05.94 & (32) & $-$33:33:33.6 & (8.0) &
 7.7 & 0.5 & $81\times49$ & 173 \\[0.15cm]
D                  &  NVSS~J191717$-$333224 & 19:17:17.84 & (52) & $-$33:32:24.4 & (5.7) &
 3.7 & 0.1 & $57\times42$ & 132 \\[0.15cm]
E                  &  GMRT163\_01\tmb       & 19:16:45.13 &      & $-$33:28:48.9 &       & 
 161 &  41 & $71\times14$ &  34 \\
\hline
\end{tabular}
\end{center}
\tablefoot{See Notes in Table \ref{t:relic}.}\\
\tablefoottext{a}{Blended in the TGSS catalog.} \\
\tablefoottext{b}{Position uncertainties from TGSS are $\sim4\arcsec$.}\\
\tablefoottext{c}{Not in the SUMSS catalog.}\\
\end{table*}

Radio relics and radio halos are diffuse, non-thermal emission features that have no obvious
connection with individual cluster galaxies and are often associated with merging activity in
massive clusters of galaxies \citep[see][for a recent review]{feretti12}. We searched for such 
features around \plck\ in the high-resolution 150~MHz images of the TIFR GMRT Sky 
Survey (TGSS)\footnote{\url{http://tgss.ncra.tifr.res.in/}} Data Release 5 and in 
VizieR\footnote{\url{http://vizier.u-strasbg.fr/viz-bin/VizieR}} \citep{ochsenbein00} for 
additional archival data.

Figure \ref{f:radio} shows the intensity map at 150~MHz from the TGSS with blue contours at
(3, 5, 7, 15)$\sigma$, where $\sigma=11.9\,\mathrm{mJy\,beam^{-1}}$ is the background 
rms level. Green and red contours show 1.4~GHz and 843~MHz emission from the NRAO VLA Sky Survey 
\citep[NVSS;][]{nvss} and the Sydney University Molonglo Sky Survey \citep[SUMSS;][]{sumss}, 
respectively. Both sets of contours are shown at (3, 5, 10, 20)$\sigma$, where 
$\sigma=0.51\,\mathrm{mJy\,beam^{-1}}$ and $2.0\,\mathrm{mJy\,beam^{-1}}$ in the NVSS and SUMSS 
images, respectively. There is significant ($>5\sigma$) emission around \plck\ in all three 
frequencies at coincident locations. Moreover, this emission is extended in the TGSS and NVSS 
images.

We identify a tangentially extended radio relic in the TGSS image, coincident with emission at the 
other frequencies, although this emission is barely resolved in SUMSS and NVSS (the extent of the 
emission is roughly 2 beams in both low-resolution images). The multi-frequency properties of this 
relic are given in Table \ref{t:relic}. Radio relics span a wide range of spectral indices, $\alpha$ 
(where $F_\nu\propto\nu^{-\alpha}$), from $\alpha\sim1$ up to $\alpha\sim3$ \citep{feretti12}. We 
give a preliminary estimate of the integrated spectral index of the relic by fitting a power-law to 
the 150~MHz flux combined with NVSS and SUMSS, one at a time. From both combinations we measure 
$0.9\lesssim\alpha\lesssim1.4$ at the 68\% level. Measuring the spectral index from all three 
frequencies gives a shallower but consistent spectral index $\alpha\sim0.7-1.1$, suggesting that the 
emission at 843 MHz and/or 1.4 GHz may be contaminated by unresolved point sources, thus boosting 
the flux and lowering $\alpha$. A spectral index measured using both 843~MHz and 1.4~GHz would be 
more affected by this contamination since these two frequencies are closer together (in log-space) 
than any of them is to the TGSS frequency.

We confirm that there are no X-ray point sources associated with any of the radio emission from 
the \xmm\ image. As with the relic, sources A, B, C and E have no counterparts in the optical 
images, nor in the Near Infrared (NIR) from the 2 Micron All Sky Survey
\citep[2MASS,][]{2mass} or the Mid Infrared (MIR) from the Wide-field Infrared Survey Explorer All 
Sky Survey \citep[WISE,][]{wise}, within their nominal position uncertainties. Source D has two
plausible counterparts from the 2MASS and WISE (merged into one source) catalogs. Both are stars,
and are also seen in our optical images. It is therefore likely that source D is a radio point
source. Given its high flux and shape in the TGSS image, source A is also likely a point 
source, or two blended point sources.

Because the relic elongation is approximately in the same direction as the TGSS beam and source E, 
we use source E (which can be regarded as noise-dominated, being much less significant and 
not detected at any other frequency) as a control for the significance of the relic accounting for 
the TGSS beam. As seen from Tables \ref{t:relic} and \ref{t:radio}, both the relic and source E 
have similar sizes and position angles. Figure \ref{f:radio} shows however that the relic is much 
more significant than source E. Moreover, the following exercise shows that in the case of source 
E, the large size is a consequence of the background noise and the beam, whereas the source we
associate to the radio relic is significantly extended over the background. We re-measured fluxes 
for these two sources on images in which we masked all pixels with values below 
$3\sigma=35.7\,\mathrm{mJy\,beam^{-1}}$. More than half the emission associated with source E comes 
from pixels with $<3\sigma$ emission, and the major axis is halved in this masked map. From the 
relic, in contrast, we still measure $\sim70\%$ of the total flux, and the major axis is 75\% of the 
size measured in the original map.

\section{Discussion \& Conclusions}\label{s:conc}

We present a multi-wavelength analysis of \plck, one of the massive galaxy clusters recently 
discovered by the \planck\ satellite using the SZ effect. Optical confirmation from GMOS imaging 
clearly shows a red sequence of galaxies with a dominant BCG, both undisputable characteristics of 
galaxy clusters. There is also a strongly lensed giant arc which is composed of two partially 
merged images of a background galaxy. Spectroscopy of 6 cluster members plus the giant arc show 
that the cluster is at $z=0.516\pm0.002$ and that the arc is at $z_{\rm arc}=1.601$. With these data 
we have performed a strong lensing analysis, confirming a third image for the source producing the 
arc. We use \lenstool\ to obtain a mass model for the cluster including the contribution from 
cluster galaxies, and estimate an Einstein mass \emass, within a median Einstein ring $r_E\simeq190$ 
kpc, corresponding to an angular size $\theta_E(z_s=1.6)\simeq30\arcsec$. Compared to the universal 
Einstein ring distribution derived by \cite{zitrin12}, \plck\ is among the 5\% strongest 
gravitational lenses in the Universe. By integrating the 3-dimensional NFW profile we estimate 
\mass. We also run \lenstool\ including a Gaussian prior for the X-ray mass estimated by 
\cite{planck_xmm} and find \mslx, marginally consistent with the mass estimated from strong lensing 
alone. The Einstein mass does not change significantly when including the X-ray constraint, because 
the latter is constrained directly by the strongly lensed galaxy. The inclusion of the X-ray mass 
constraint does help to exclude a high-mass, low-concentration solution which is allowed by the 
strong lensing-only model.

Examination of archival high-resolution radio data from the TIFR GMRT Sky Survey at 150~MHz reveals 
the presence of a radio relic at approximately 250 kpc from the cluster center. Significant emission 
is also detected in low-resolution images from NVSS at 1.4~GHz and SUMSS at 843~MHz. A preliminary 
measurement of the integrated spectral index yields $\alpha\sim0.9-1.4$. We find no detectable point 
sources contributing significantly to the radio emission in the \xmm\ or Gemini images, nor from 
archival observations in the NIR or MIR. This radio emission likely originated from recent 
merging activity, but the available data do not allow for a detailed study of possible merging 
events. The origin of the radio emission will be addressed with future observations.

\begin{acknowledgements}
We thank Timo Anguita for help with the arc spectrum.
CS acknowledges support from the European Research Council under FP7 grant number 279396 awarded to
H.\ Hoekstra.
CS and LFB have been supported by ``Centro de Astrof\'isica FONDAP'' 15010003, Centro BASAL-CATA,
by FONDECYT under project 1120676 and by ALMA-CONICYT under projects 31090002 and 31100003.
JPH acknowledges support from NASA ADAP grant number NNX11AJ48G.

This research has made use of the VizieR catalogue access tool, CDS, Strasbourg, France. The
original description of the VizieR service was published in A\&AS 143, 23.

This research work has used the TIFR GMRT Sky Survey (\url{http://tgss.ncra.tifr.res.in}) data 
products.
We thank the staff of the GMRT who have made these observations possible. GMRT is run by the 
National Centre for Radio Astrophysics of the Tata Institute of Fundamental Research.

All plots in this paper were generated with the python package matplotlib \citep{hunter07}.

\end{acknowledgements}

\end{document}